\documentclass[traditabstract]{aa}
\usepackage{longtable}
\usepackage{comment}
\usepackage{journals}
\usepackage{placeins}
\usepackage{textcomp}
\usepackage{color}
\usepackage{fixltx2e}
\usepackage{graphicx,natbib,txfonts,color}
\bibpunct{(}{)}{;}{a}{}{,}

\begin{document}

\title{MAIA, a three-channel imager for asteroseismology: \\instrument design
\thanks{Based on observations made with the Mercator Telescope,
  operated on the island of La Palma by the Flemish Community, at the
  Spanish Observatorio del Roque de los Muchachos of the Instituto de
  Astrof\'{i}sica de Canarias. }
}

\author{Gert Raskin\inst{1}
\and Steven Bloemen\inst{1,4}
\and Johan Morren\inst{1}
\and Jesus Perez Padilla\inst{1,2}
\and Saskia Prins\inst{1,2}
\and Wim Pessemier\inst{1}
\and Jeroen Vandersteen\inst{3}
\and Florian Merges\inst{1,2}
\and Roy \O{}stensen \inst{1}
\and Hans Van Winckel\inst{1}
\and Conny Aerts\inst{1,4}
}

\institute{
Institute of Astronomy, Department of Physics and Astronomy, 
KU\,Leuven, Celestijnenlaan 200D,
B-3001 Leuven, Belgium 
\and 
Mercator Telescope, 
Roque de los Muchachos Observatory,
La Palma, Spain
\and
RHEA, European Space Agency -- ESTEC, Keplerlaan 1, 2201 AZ Noordwijk, 
The Netherlands
\and
Department of Astrophysics, IMAPP, Radboud University 
Nijmegen, PO Box 9010, 6500 GL Nijmegen, The Netherlands}
\date{Received / Accepted }

\offprints{Gert.Raskin@ster.kuleuven.be}

\authorrunning{G. Raskin et al.}
\titlerunning{MAIA, a three-channel imager for Asteroseismology}

\abstract{ MAIA, an acronym for Mercator Advanced Imager for Asteroseismology,
  is a three-channel instrument that targets fast-cadence three-colour
  photometry, installed at the 1.2-m Mercator telescope at the Roque de los
  Muchachos at La Palma (Canary Islands, Spain). This instrument observes a
  9.4\,x\,14.1\,arcmin$^2$ Field-of-View simultaneously in three wavelength
  bands on three large frame-transfer CCDs. These
  detectors were developed for ESA's cancelled Eddington space mission and were
  offered on permanent loan to the Institute of Astronomy (KU\,Leuven,
  Belgium).  MAIA uses its own $ugr$ photometric system that is a crude approximation of the SDSS system.
 The instrument is designed to perform multi-colour observations for
  asteroseismology, with specific emphasis on subdwarf and white dwarf single
  and binary stars.  We describe the design of the instrument, discuss key
  components, and report on its performance and first results.
}

\keywords{Asteroseismology --
Instrumentation: photometers --
Instrumentation: detectors --
Techniques: photometric --
Stars: Variables
}
\maketitle

\section{Introduction}\label{sect:intro}

The MAIA project found its origin in the Eddington space mission that was
cancelled by the European Space Agency (ESA) in 2003. The main objectives of
Eddington were asteroseismology and exoplanet transit detections
\citep{Favata2000}. A number of charge-coupled device (CCD) detectors had been
developed and procured specifically for this mission, based on the well-known
e2v CCD42-xx series. These are large frame-transfer devices, designated as
CCD42-C0.

After the cancellation of the Eddington mission, some of these detectors were
offered to the astronomical community provided that they would be used for the
original science goals of the cancelled Eddington mission. One of us
(CA) defined a proposal to use the CCDs to perform asteroseismic studies of
evolved stars, in particular of subdwarf B (sdB) stars, by constructing a camera
to be installed at the 1.2-m Mercator telescope. As such, the Institute of
Astronomy of the University of Leuven (KU\,Leuven, Belgium) received the
Eddington detectors through a permanent loan agreement between ESA and
KU\,Leuven.  The present paper presents the scientific specifications, design,
installation and the first performance tests of the Mercator Advanced Imager for
Asteroseismology (MAIA).

The Mercator telescope \citep{raskin04} is a modern, semi-robotic 1.2-m
telescope, installed at the Roque de los Muchachos Observatory on La Palma
(Canary Islands, Spain), funded by the Flemish community of Belgium, and
operated by the KU\,Leuven Institute of Astronomy. This telescope provides
Belgian astronomers with permanent access to an intermediate-size telescope,
opening up the niche of high-precision long-term studies of time-variable
astronomical phenomena. This has allowed the Leuven team and its collaborators
to continue its leading role in the field of variable single and
multiple star research, where long time-series are essential. 
The operational model of Mercator is focused on and committed to long-term
programmes which are particularly important for the study of stellar pulsations,
where the time span of the observations defines the precision with which
independent pulsation frequencies can be detected and identified. Another prime
science case for the Mercator telescope is the study of binary stars, with
specific emphasis on long-period evolved binaries where orbits of one
to several years are commonly found.

Currently, the Mercator telescope is also equipped with a stable and efficient
high-resolution \'echelle spectrograph \citep[HERMES,][]{Raskin11}. The MEROPE
imager \citep{davignon04} was already equipped with one of the Eddington detectors from
2009 until 2012 for performance tests and characterisation purposes in preparation for MAIA \citep[MEROPE
  II,][]{oestensen10}, and has since been decommissioned in order to install
MAIA.

The preliminary design of MAIA was first presented in \citet{Vandersteen10}. In
the mean time, we have built the instrument, installed it on the telescope
(2012) and almost completed the commissioning (to be finished by the end of 2013).  In this
paper, we discuss the scientific motivation behind the project and its
consequences for the instrument requirements, as well as a detailed description
of its design.  We also report on the measured performance of MAIA and present
some first on-sky results. In a subsequent paper (Bloemen et al., in
preparation), we plan to present the detailed commissioning of the instrument, the data reduction software necessary for the
optimal scientific use of the instrument, and illustrate its capacity for
asteroseismology of pulsating subdwarf B stars.

\section{Science case}
\label{science}

Putting efforts into ground-based asteroseismology remains very relevant for
particular types of stars, even if the space missions MOST \citep{walker2003},
CoRoT \citep{auvergne2009} and {\it Kepler\/} \citep{gilliland2010} have resulted in a revolution in our knowledge of stellar interiors.
Indeed, while the $\mu$mag precision white-light space photometry led to various
recent breakthroughs to improve stellar physics, e.g., the ability to deduce
inhomogeneously mixed zones around the convective core of massive stars 
\citep[e.g.][]{Degroote10}, the connection between beating of oscillation modes and
outbursts in Be stars \citep[e.g.][]{huat2009}, the discovery of
gravito-inertial modes in fast rotators \citep[e.g.][]{neiner2012,papics2012}, the power to discriminate between hydrogen-shell and core-helium
burning red giants from their dipole mixed modes \citep[e.g.][]{bedding2011},
the derivation of core to envelope rotation of subgiants \citep{deheuvels2012}
and of red giants \citep{beck2012}, to list just a few, some types
of pulsators could not be studied, or insufficiently so, by these missions. This
is particularly the case for ultra-fast pulsators, such as the roAp stars,
pressure-mode pulsating subdwarfs and gravity-mode pulsating white dwarfs, with
pulsation periods of the order of one to a few minutes and of which too few class
members occurred in the Field-of-View (FoV) of the satellites to do ensemble
asteroseismology of those classes. MAIA was designed to be able to target
asteroseismology of such fast and relatively faint pulsators.

A particular point of attention, bridging two important scientific aims of the
Mercator telescope which cannot be done with a space missions alone, was to
build an instrument to complement the HERMES spectrograph for the study of
binary pulsators that underwent a phase of common envelope or stable Roche-lobe
overflow during the red-giant branch. Presently, the physics of the common
envelope phase is still described by an ad-hoc parameter connected with the
envelope ejection efficiency. In this model, it is assumed that the orbital
energy released during the spiral-in is used to eject the common envelope of the
two stars and results in a short-period binary in the core-helium burning
phase.  In the case of stable Roche-lobe overflow, one expects to find
long-period horizontal branch binaries but hardly any have been found so far due
to the long-term monitoring requirement \citep[e.g.][]{vos2012}.  Luckily,
pulsators on the horizontal branch that have passed these poorly known
evolutionary stages exist and these allow seismic tuning of their envelope
structure and mass, and subsequent backtracking of their previous evolution
\citep[e.g.][]{hu2008}. Despite impressive results on horizontal branch star
asteroseismology from {\it Kepler\/} 
\citep[e.g.][]{vangrootel2010,charpinet2011,reed2011,oestensen2013}, only a few {\it
  Kepler\/} targets are suitable to tackle the binary evolution case \citep{oestensen10b,pablo2012}.

In general, asteroseismology allows one to constrain the internal structure of stars
by means of forward modelling, starting from a frequency analysis and secure
identification of the spherical degree $\ell$ of several detected pulsation
modes. Such forward modelling delivers the stellar fundamental parameters with
typically an order of magnitude better precision than can be achieved from
the comparison of classical data, such as spectrum analysis and/or
interferometry, with stellar models. Moreover, asteroseismology allows, in
principle, to improve the input physics of the current stellar models by
exploiting the seismic information in full details, as has been achieved for
the Sun \citep[e.g.][]{christensen2002}.  

A prerequisite for a successful seismic application, is an unambiguous
identification of the detected pulsation modes.  Three well established methods
for identifying the modes of a pulsating star are available:
\begin{enumerate}
\item
from white light or single band photometry by means of recognition of frequency
or period spacings for modes of consecutive radial orders, or from rotationally
split multiplets;
\item
from multi-band photometry, employing simultaneous observations in at least
three different wavelength bands, through interpretation of the measured
amplitude ratios and/or phase differences;
\item
from time-resolved high signal-to-noise spectroscopy allowing the
line-profile variations of unblended spectral lines to be interpreted
\end{enumerate}
\citep[e.g.][Chapter~6]{aerts2010c}. MAIA is designed to use method 2 in the
case of relatively faint and fast pulsators with amplitudes near 100\,$\mu$mag or
higher, whose binarity can be modelled from long-term HERMES
spectroscopy. However, since this particular science case is more demanding than
seismic applications to most other classes of pulsators with similar amplitudes,
it can also treat various other cases, and in particular stars where space
asteroseismology was not possible or very limited so far, e.g., hot massive
supergiant asteroseismology.

\section{Instrument requirements}
\label{sec:requirements}
The following high-level instrument requirements could be derived from the science case that was presented in the previous section:
\vspace{1mm}

\newcounter{req}\stepcounter{req}
\noindent{\textit{\arabic{req}.\,Simultaneous measurements in at least three optical wavelength
bands.}}
\stepcounter{req}
To increase the contrast of the amplitude ratios, the separation between the dichroic cut points should be as wide as possible. 
\vspace{1mm}

\noindent{\textit{\arabic{req}.\,High sample rate.}}
\stepcounter{req}
For accurate sampling of the shortest pulsation periods, cycle times as short as a few seconds are required. Obtaining acceptable exposure times in such short periods means that the dead time between exposures should be negligible.
\vspace{1mm}

\noindent{\textit{\arabic{req}.~Uninterrupted long-term availability.}}
\stepcounter{req}
The exploitation scheme of the Mercator telescope foresees to have all instruments permanently installed. Switching between instruments only takes a few minutes. Consequently, MAIA is almost permanently available for observations. This is essential for obtaining accurate oscillation frequencies as they require continuous time series of observations with a long time base.
\vspace{1mm}

\noindent{\textit{\arabic{req}.\,Wide field of view.}}
\stepcounter{req}
Precise measurements of the brightness variation of pulsating stars require
the simultaneous observation of preferably several reference stars. Especially
in the $u$ band, it is difficult to find bright enough comparison stars. To increase the probability of finding sufficiently bright reference
sources, MAIA should cover a large field of view. The MEROPE II camera
on the Mercator telescope had a FoV of 9\,x\,6.6\,arcmin$^2$. Our experience
with  this instrument showed that MAIA should cover a FoV that is at least twice
as large.
\vspace{1mm}

\noindent{\textit{\arabic{req}.\,Spatial resolution.}}
\stepcounter{req}
MAIA image quality should only be a negligible contributor to the point spread function (PSF), defined by the size of the seeing disk  in case of excellent atmospheric conditions at the Roque De Los Muchachos observatory on La Palma ($\sim$\,0.6\,arcsec).  The sampling of an 0.6-arcsec PSF should satisfy the Nyquist criterion. 
\vspace{1mm}

\noindent{\textit{\arabic{req}.\,High throughput.}}
\stepcounter{req}
Obviously, high sample rates go hand in hand with short exposure times. However,
many possible MAIA targets are faint or span a broad brightness range. This
is especially relevant when considering the different wavelength ranges that
are sampled simultaneously. For a typical reference star, the $u$-band flux can
be very small and many magnitudes fainter than the $r$ or $g$-band flux. Given the limited collecting  area of the 1.2-m Mercator telescope, it is of high importance to reduce to an
absolute minimum the number of photons that get lost on their way through the instrument.
\vspace{1mm}

\noindent{\textit{\arabic{req}.\,Portability.}}
\stepcounter{req}
At the time of writing, no concrete plans exist to take MAIA to other
telescopes than Mercator. Nevertheless, in order to extend MAIA's scope
towards fainter targets, we also consider deploying the instrument at larger
aperture telescopes in the future. Therefore, MAIA should be designed with
portability in mind and the optics should be adaptable to a telescope that is
at least twice as large as the 1.2-m Mercator telescope.
\vspace{1mm}

\noindent{\textit{\arabic{req}.\,Detectors.}}
\stepcounter{req}
The budget for building MAIA leaves no room to procure dedicated detectors and hence
relies completely on the use of the Eddington CCD42-C0 devices provided by ESA.
Therefore, the instrument design has to be adapted to the properties of these
detectors. Fortunately, the pixel size and the pixel count of the Eddington CCDs are well suited for the type of observations targeted  by MAIA (see section~\ref{sec:CCD-42C0}).

\section{Instrument design}
\label{sec:design}
The design of MAIA is loosely inspired by ULTRACAM
\citep{Dhillon2007} but in contrast to this instrument, MAIA does not
target sub-second sample rates. Due to the much larger detectors, however, MAIA
offers a much larger FoV. In the following sections, we present the most
important design aspects of MAIA.

\subsection{Detector system}
\label{sec:detector}

\subsubsection{
\label{sec:CCD-42C0}
The CCD42-C0 frame-transfer detectors}
\begin{figure}
\centering
\resizebox{\hsize}{!}{\includegraphics[angle=270]{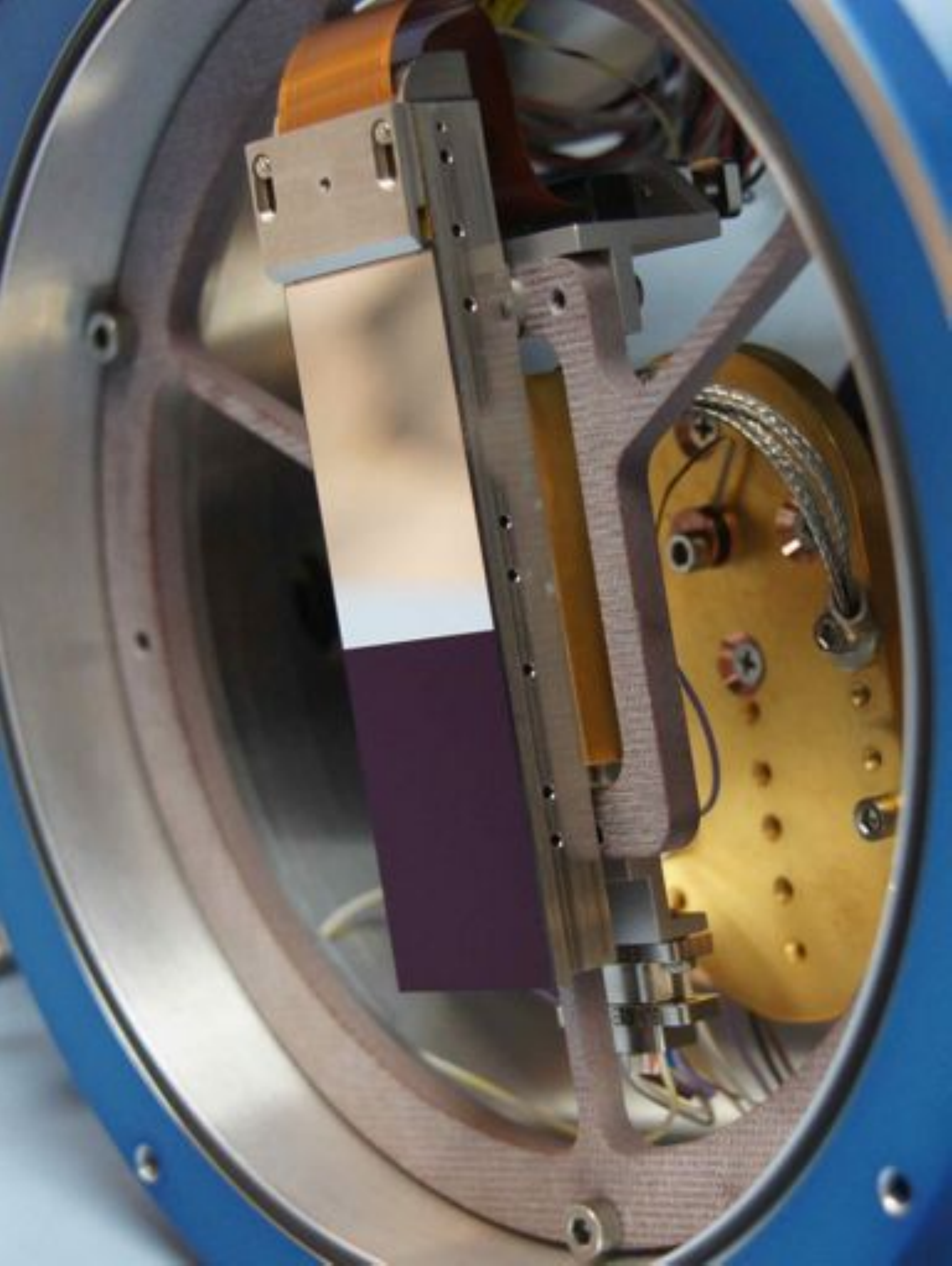}}
\caption{\label{fig:detector}
Picture of a CCD42-C0 frame-transfer detector, mounted in the $u$ cryostat.}
\end{figure}

The starting point of the MAIA design are the CCD42-C0 detectors that
were developed by e2v (UK) for the Eddington space mission
\citep{lumb03}. These are thinned back-illuminated frame-transfer (FT)
devices with a basic mid-band anti-reflection coating.  Unfortunately,
this coating has a rather poor UV response. An enhanced broad-band
coating with higher throughput at short wavelengths would have been a
much better match for the important MAIA $u$\,channel. As a result, the UV quantum efficiency (QE) is much lower than expected (less than 20\% for wavelengths below 370\,nm, see figure~\ref{fig:transmission}).
The chips have
a format of 2048\,x\,6144 13.5-$\mu$m pixels, split in halves between
a 2k\,x\,3k imaging area, where charge collection takes place, and an
equally large storage area. The effective area of these unique
detectors is substantially larger than that of any other commercially
available FT device. Figure~\ref{fig:detector} shows a picture of a
CCD42-C0 detector, mounted on a fibre-glass spider in its
cryostat. The dark part at the left-hand side is the
photo-sensitive imaging area and the shiny part at the right is the
aluminium-covered storage area. The detector is very long
($\sim$\,110\,mm) so it is installed with the imaging area out of
centre, in order to reduce the diameter and volume of the cryostat. Consequently, the window in the cryostat is similarly decentred. 

The FT design allows a rapid shift of the image data from the imaging to the storage
area, so that integration can proceed in the imaging area while the previous
image is being digitised from the storage area.
The frame
transfer shift takes only 295\,ms, while the image data readout of a full
frame takes between 43 and 30 seconds, depending on the controller digitisation speed settings
and the read-out noise requirements (see table\,\ref{tab:readout}).
The time required to transfer a full-frame image from the imaging 
to the storage area is more than 100 times smaller than the full-frame read-out time. The dead time during
time-series observations is only limited to this 295\,ms frame transfer.
The usage of the FT mode implies that the
integration time should be at least as long as the duration of the image
read-out from the store area.  When shorter integration times are required,
data can also be acquired using a classic read-out mode with a destructive
clear prior to the integration, but in that case the full read-out overhead applies.

All three MAIA cameras are equipped with a mechanical, individually
controlled, iris shutter that allows the detectors to be used in
classic (non-FT) mode. In FT mode the shutters are continuously open.

\subsubsection{Detector cooling}
\label{sect:cooling}
Although MAIA's main science case only needs short integration times,
instrument versatility also required the possibility of obtaining long
exposures. Even when observing the faintest targets, detector dark
current should remain a negligible noise source. This means that it
should be significantly smaller than the flux from the new-moon sky,
which can be as small as 0.1\,e$^-$\,pixel$^{-1}$s$^{-1}$ in the $u$
band for MAIA on the Mercator telescope. In this case, the dark
current should not exceed
0.05\,e$^-$\,pixel$^{-1}$s$^{-1}$. Laboratory measurements show that
the CCD42-C0 needs a temperature below 190\,K to reach this dark
current level. Alternatively, operating the detector in inverted state
would reduce the dark current at a much higher temperature. However, a
negative consequence of this inversion is a substantially reduced
full-well capacity.  This seriously limits the dynamic range of the
cameras so we discarded this solution.

We explored various technologies to cool the MAIA detectors to
190\,K. Thermo-electric cooling of very large detectors like the
CCD42-C0 by multi-stage Peltier elements, turned out to be extremely
difficult. Moreover, in the absence of cryogenic temperatures,
cryo-pumping could not be used to maintain a long-term vacuum in the
cryostat.  Liquid nitrogen (LN$_{2}$) and Joule-Thompsom (JT) cooling
with e.g. a CryoTiger system are long-time proven solutions. However, the
use of LN$_2$ was discarded because of the burden of filling three cryostats
at the start and end of each night. JT coolers
for three cryostats implied two or three compressors and six gas lines with 
heavy stainless steel braiding, 
a very bulky system to accommodate at the Nasmyth
focal station of Mercator. We therefore looked for an alternative and
settled on the use of compact free-piston Stirling coolers to cool the
MAIA detectors. After some market research, we selected the MT CryoTel
cryocooler (Sunpower, USA) that can evacuate 5\,W of heat at 77\,K with an
electrical input power of 80\,W \citep{unger04}. These are very
convenient and compact devices with a weight of only 2.1\,kg. The
vibrations generated by the linear compressor are the only important
disadvantage.  In section~\ref{sec:vibrations} we discuss how these
vibrations are reduced to an acceptable level.

The thermal link between the detector and the cold head of the cooler is dimensioned to obtain a detector equilibrium temperature around 155\,K. A resistive heater in closed loop heats the detector and stabilises its temperature at 165\,K with an accuracy of a few 0.01\,K. We use a 
programmable logic controller (PLC) and industrial hardware to implement this
feedback loop, as well as for all other instrument control tasks
\citep{pessemier2012}. A human-machine interface (HMI) with touch screen is
mounted on the instrument.
It provides a visual indication of the status (e.g. detector temperature, cooling power, etc.) and allows manual technical control of MAIA (e.g.
changing temperature setpoint, starting a controlled warm-up procedure, etc.).

\subsubsection{Data acquisition}
\label{sec:daq}
The three MAIA detectors are read out and controlled by a standard
generation-III SDSU detector controller (Astronomical Research Cameras,
USA) \citep{Leach00}. Volume and weight restrictions required the use of the
small 6-slot housing and small power supply. Therefore, the controller could
only be equipped with two dual channel video boards, and thus only four read
ports can be used. Hence, although the detectors have a dual-port split
serial register, only single-port read-out is currently implemented. Windowing
and binning are used to reduce the read time if short exposure
times are required. 

We have implemented standard single windowing of the detectors as well
as a multi-window mode.  In the multi-window mode, up to ten windows
can be defined anywhere on the detector. All windows must have the
same dimensions. The choice of window height and vertical position is free 
but the windows should not overlap.
We found that serial skips between prescan, science windows and overscan, 
causes a disturbing bias level gradient of several digitisation units at the horizontal start of each window. 
Therefore, we fixed the horizontal extent of the windows to
the full width of the detector, including prescan and overscan (2150
pixels). 

In table~\ref{tab:readout}, characteristics of the two main detector
read-out modes are given. In case the detectors are operated in frame-transfer
mode, the minimum exposure time is defined by the read-out time. When
combining the frame-transfer mode with windowing of the detector, the
single-row read-out time can be used to determine the read-out and
minimum exposure times by multiplying with the number of window
rows. As an example, two windows of 60 rows each (one placed at the top and one at the bottom part of the detector, both spanning the entire detector width), require a minimum cycle time of two seconds in the most common \textit{Slow} read-out mode . 

\begin{table}
\centering
\caption{\label{tab:readout}
MAIA detector read-out modes. Read noise is slightly lower in $u$ and $r$ than it is in $g$.}
\label{tab:readout}
\begin{tabular}{ccc}
\hline\hline\noalign{\smallskip} \rule[0mm]{0mm}{1mm}
Read-out mode & \textit{Slow} & \textit{Fast} \\
\noalign{\smallskip}\hline\noalign{\smallskip}
Read-out frequency & 152~kPixel\,s$^{-1}$ & 219~kPixel\,s$^{-1}$\\
Read-out noise & 3.5\,--\,4\,e$^-$ & 4.5\,--\,5\,e$^-$\\
Single row read-out time & 14.1\,ms & 9.8\,ms\\
Full frame read-out time & 43.5\,s & 30.3\,s \\
Full frame transfer time & 295\,ms & 295\,ms\\
Conversion gain & 0.8~ADU\,/\,e$^-$ & 0.4~ADU\,/\,e$^-$\\
\hline
\end{tabular}
\end{table}

A combination of the lower instrumental efficiency in the \textit{u} band,
the often smaller fluxes due to the intrinsic energy distribution of
the sources and the high extinction of the earth atmosphere at blue wavelengths, means that
saturation can occur in the $g$ or $r$ channel before a
sufficiently high signal-to-noise ratio is obtained in \textit{u}.  Therefore, we modified the code for the timing-board digital
signal processor (DSP) to allow different integration times
for the individual channels.  
After defining a nominal integration time that
needs to be longer than the minimal exposure time, each detector can
integrate for 1x, 2x, or 4x the nominal integration time.  
Read-out from the storage area occurs simultaneously for
all detectors at every nominal integration time, but image data are only
shifted into the storage area at the defined multiples of the integration
times.

Currently, all frames obtain a time stamp, based on a computer clock that is synchronised through 
Network Time Protocol (NTP) with the stratum 2 time server of the observatory.  This time stamp may 
suffer from several variable delays up to 100\,ms due to varying load on the network and the 
data-acquisition host. Because this is not the most reliable way 
to ensure accurate timing, we plan to upgrade the Mercator telescope control 
system with a directly-connected GPS time server in the near future. This time server will also be linked 
to the MAIA  detector controller in order to synchronise the exposures and to deliver reliable and accurate 
time stamps.

\subsection{Optical design}
\label{sec:optics}

\subsubsection{Layout}
\label{sec:layout}
\begin{figure}
\centering
\resizebox{\hsize}{!}{\includegraphics*{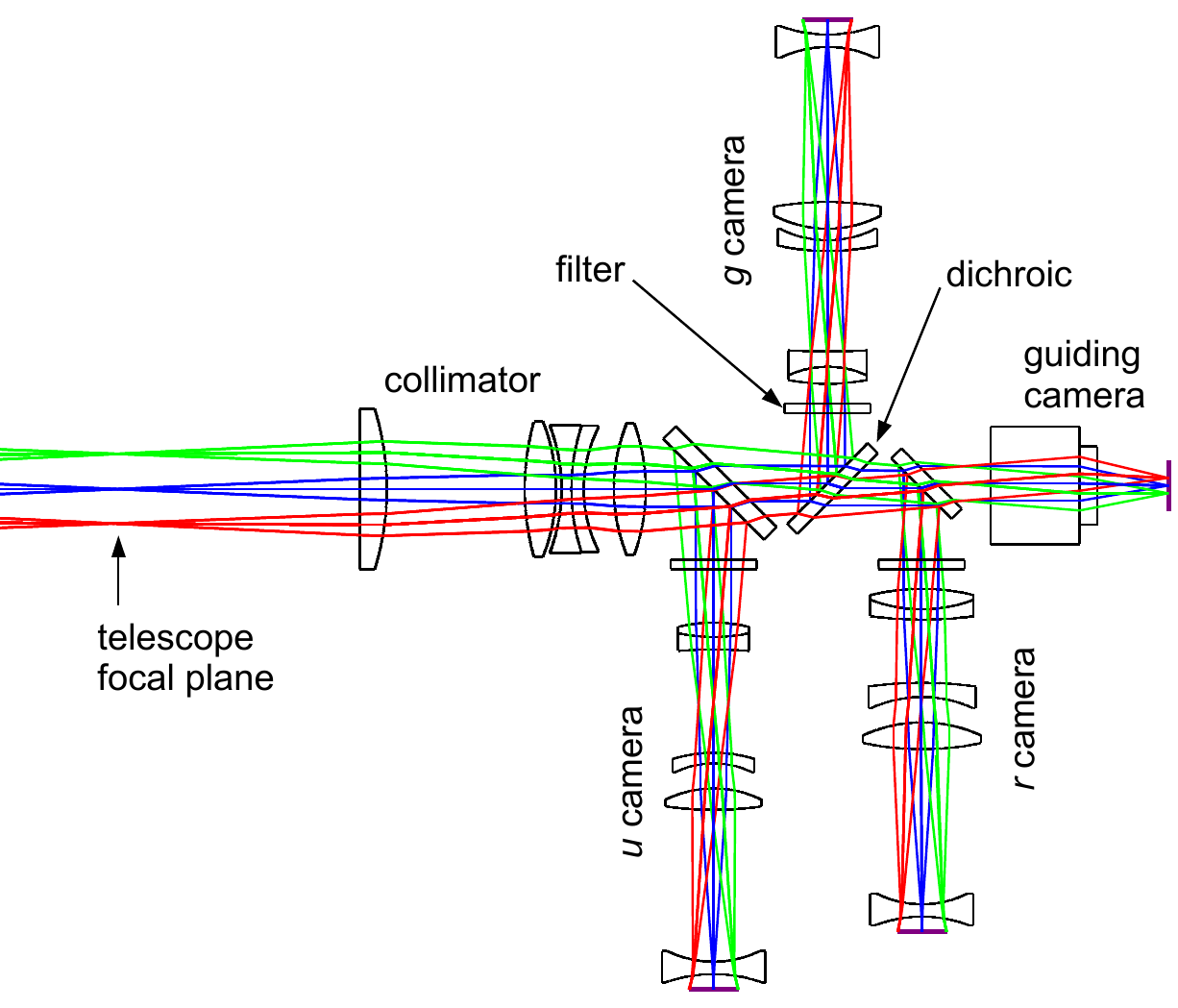}}
\caption{\label{fig:raytrace}
MAIA optical layout and ray trace. The $r$ channel orientation is perpendicular to the collimator-$u$-$g$ plane but  for clarity, it is 90$^{\circ}$ rotated in the drawing.}
\end{figure}

A ray trace, illustrating the operation of the MAIA optics, is given in
figure~\ref{fig:raytrace}. The light coming from the telescope is first collimated
and then split in different wavelength bands by three dichroic beam splitters.
Each dichroic splits off one of the three science beams. Three cameras re-image the separated
beams on the corresponding detectors. Additional filters can be inserted between
the dichroics and cameras for a better definition of the pass bands. Only the reddest light passes through all three
dichroics and is lost for science measurements. This near-infrared part
of the beam ($\lambda>875$\,nm) is collected by the MAIA guiding camera to be used
for on-axis telescope guiding. As the Mercator telescope does not possess a 
guiding facility, each instruments needs to be equipped with an integrated
guiding camera.

\subsubsection{Dichroics and filters}
\label{sec:dichroics}
\begin{figure*}
\centering
\includegraphics[width=18cm]{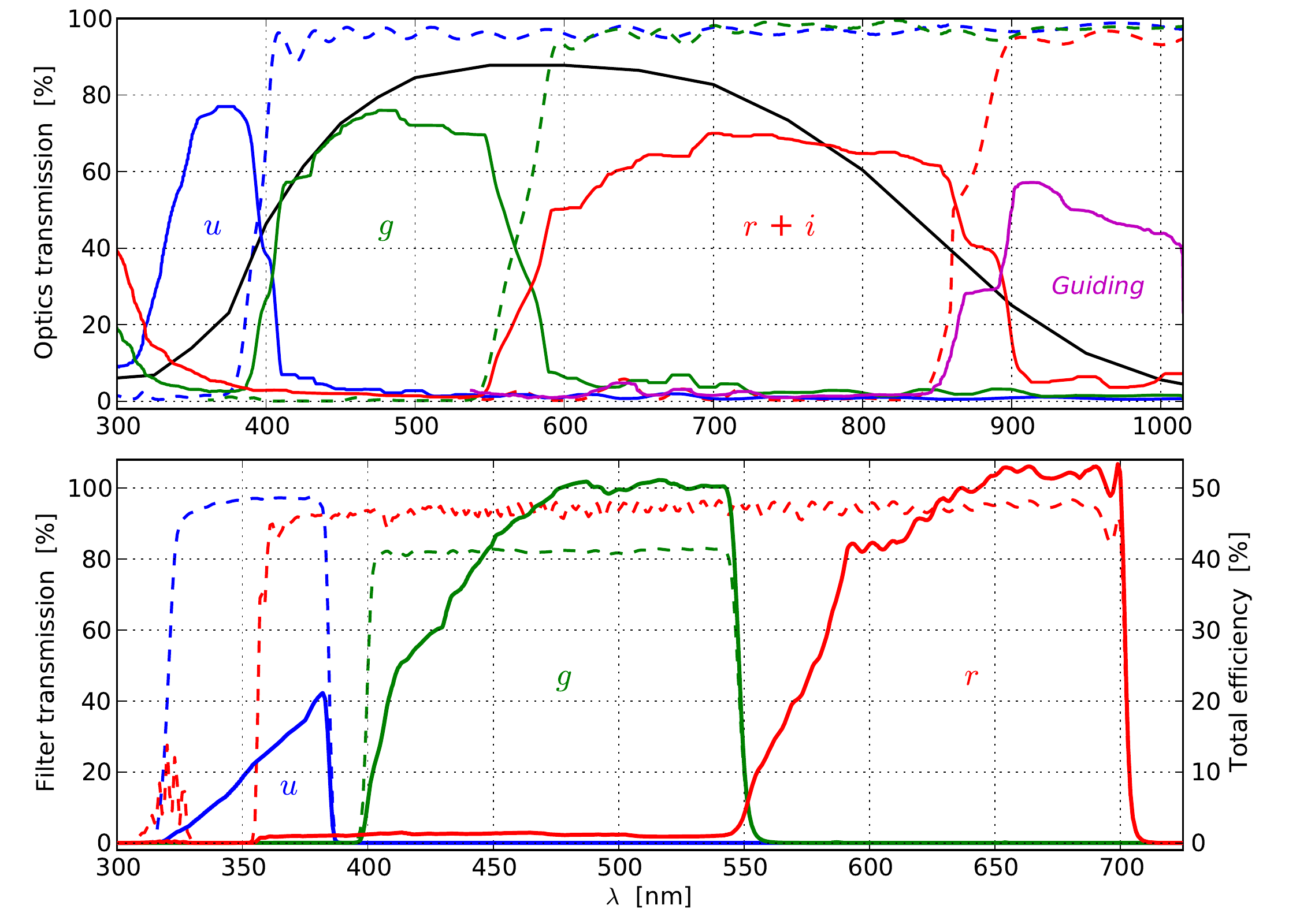}
\caption{\label{fig:transmission}
Top: detector QE (solid black curve), dichroic transmission (dashed colours) and total optics transmission (convolution of measured dichroic curves and calculated transmission of glass and AR coatings, solid colours). Bottom: filter transmission (dashed) and total calculated instrument efficiency (solid), including filters and detectors but excluding telescope and atmosphere.}
\end{figure*}

The dichroic beam splitters consist of a flat fused-silica substrate with a
dielectric coating that transmits the wavelengths longer than the cut-off
point and reflects those that are shorter. For efficiency reasons, the coating
is applied to the front surface of the beam splitters, while the rear surface is
treated with a 1\% anti-reflection coating. Originally, it was foreseen that
the dichroic cut-points would mimic the SDSS photometric system. However, due
to a manufacturing problem, the cut-off wavelengths turned out to be slightly longer than
foreseen. Moreover, due to the low detector QE in $u$, MAIA  has much lower throughput in this band than the SDSS system.  The measured 50\% cut points for unpolarised light are given in table~\ref{tab:filters}. Figure~\ref{fig:transmission}~(top) shows the measured transmission curves of the dichroics, together with the  transmission of  the other optical elements as specified. The dichroics were procured from CVI\,--\,Melles~Griot (Isle of Man). 

Dichroic beam splitters are never perfect components, and a few percent of the
flux might end up in the wrong channel. Especially the red light that gets
reflected toward the \textit{u} camera  can greatly compromise accuracy.
Therefore, additional band-pass filters follow each dichroic. SDSS $u'$ and $g'$ filters are installed by default in the $u$ and $g$ channels. The MAIA \textit{r} filter is a short-pass filter that cuts off at 700\,nm,
substantially below the dichroic upper cut point of 875\,nm. This way, we avoid that strong telluric line fluctuations compromise photometric precision.  All filters block the wavelengths outside the pass band to  less than 0.01\%.
Figure~\ref{fig:transmission}~(bottom) shows the measured transmission of the filters, as well as the total throughput of MAIA, including all optical elements and  the detectors.

Filter holders for installing different filters are available. All MAIA filters have a 
50.0\,--\,50.8\,mm diameter and a maximum thickness of 6\,mm. The filter holders
need manual insertion in the instrument and therefore, it is not recommended to change filters during an observing night. As the filters are installed in a collimated beam, their absence or different thickness will not affect the focus of the cameras.

\begin{table}
\centering
\caption{\label{tab:filters}
MAIA dichroic 50\% cut points and effective central wavelength of each filtered pass band.}
\begin{tabular}{ccc}
\hline\hline\noalign{\smallskip} \rule[0mm]{0mm}{1mm}
Band & $\lambda_{\hbox{\tiny{dichroic\,cut}}}$ & $\lambda_{\hbox{\tiny{central}}}$\\
\noalign{\smallskip}\hline\noalign{\smallskip}
\textbf{\textit{u}} & 395\,nm & 364\,nm  \\
\textbf{\textit{g}} & 574\,nm & 483\,nm\\
\textbf{\textit{r}} & 875\,nm & 632\,nm\\
\hline
\end{tabular}
\end{table}

\subsubsection{Collimator and cameras}
\label{sec:dioptrics}

To obtain a sufficiently large FoV on the CCD42-C0 detectors, the
collimator--camera combination needs to provide some focal
reduction. For a FoV of 14.1\,x\,9.4\,arcmin$^2$, a ratio of 1.44 is
required between the collimator focal length and the camera focal
length.  The focal length of the collimator was set at
\textit{f}\,=\,230\,mm, a compromise between limited overall
dimensions, requiring short focal length, and limited field angles in
collimated space, leading to small optical components but requiring
long focal length. This gives a focal length of 160\,mm for the
cameras. The plate scale at the camera focal plane is then
0.276\,arcsec\,pixel$^{-1}$, providing sufficient sampling under good
seeing conditions.

The collimator is a 5 lens system (4 singlets and 1 doublet) with an
\textit{f}/12 focal ratio, corresponding to the Mercator telescope
optics. The design was driven by having an exit-pupil quite far behind
the last collimator lens. This locates the pupil close or inside the
cameras, simplifying the camera design and greatly reducing the size
of dichroics and camera lenses. The collimator has not been designed
to operate as an independent system, but the quality of the collimated
beam is more than sufficient to avoid ghost-image issues that would
result from a non-parallel beam passing through the thick 45$^{\circ}$
dichroics. Instead, optimisation was done on the complete
telescope--collimator--camera combination. This allowed us to benefit
from additional degrees of freedom without having the image quality of
the collimator as a constraint.

All three cameras have a similar design consisting of one doublet and
three singlet lenses. The last lens is a field flattener that corrects
the combined field curvature from telescope, collimator and
camera. The field lens also acts as the vacuum seal of the
cryostat. This avoids the need for an additional plane vacuum window
and thus reduces Fresnel reflections. To limit instrument dimensions,
the three cameras do not have identical pupil locations. For the $u$
channel the pupil sits almost in the centre of the camera, the $g$
pupil lies on the first camera lens, and $r$ has the pupil on the
dichroic in front of the camera. The camera focal length of 160\,mm
corresponds with a slow \textit{f}/8.3 focal ratio. The camera
apertures have been oversized by about 33\% to decrease the FoV
reduction when using MAIA on telescopes larger than the Mercator
telescope. Nevertheless, due to the favourable location of the exit
pupil of the collimator, the diameters of the camera lenses do not
become larger than about 50\,mm. All MAIA lenses were manufactured and
coated by Optique Fichou (France). The dimensions of the dichroic beam
splitters are also adapted to the increased camera aperture. Deploying
MAIA at a larger telescope, only requires the replacement of the
collimator optics and the mechanical telescope interface ring.

\begin{figure}
\centering
\resizebox{\hsize}{!}{\includegraphics*{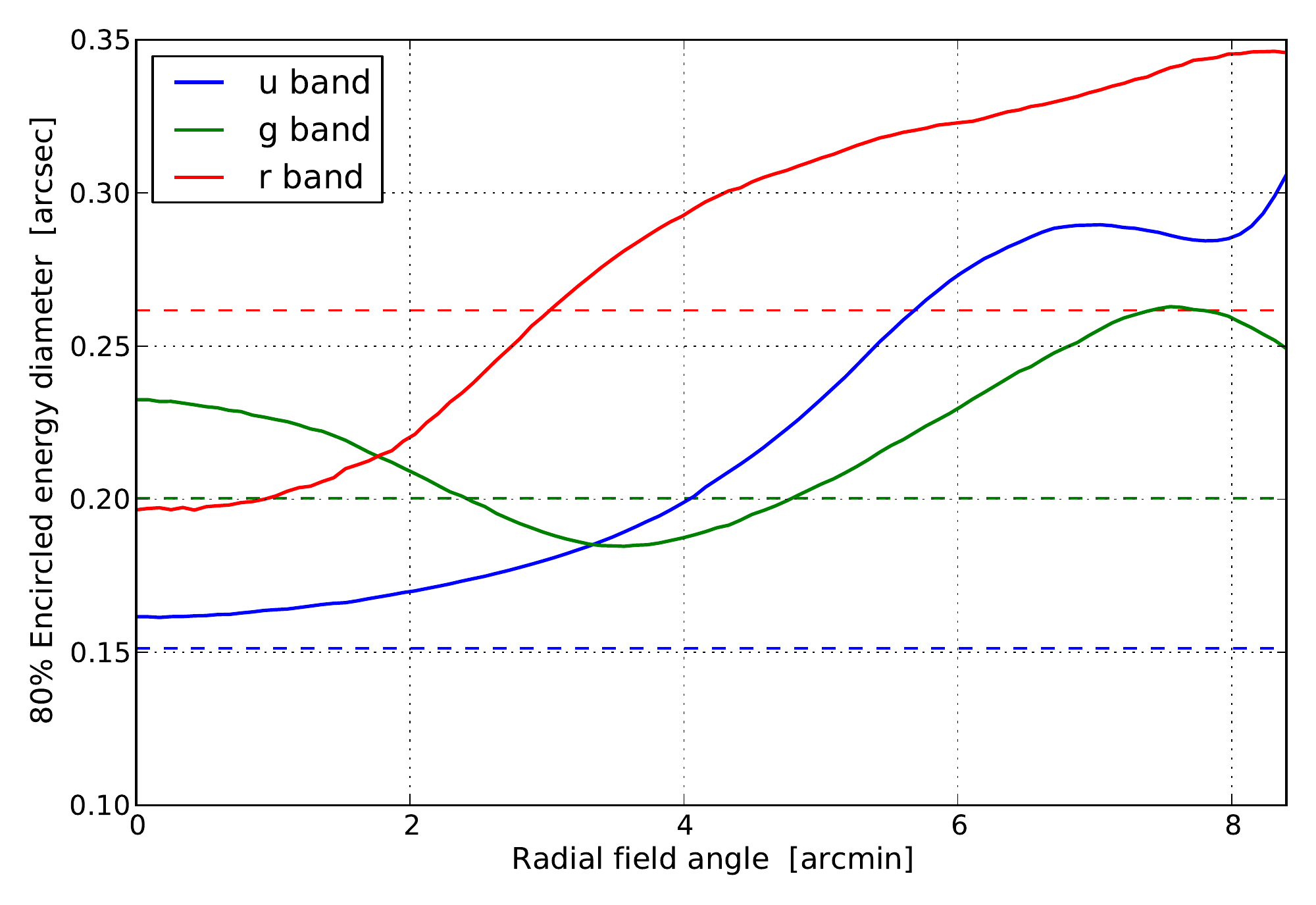}}
\caption{\label{fig:ee80}
Calculated 80\% encircled energy diameters over the FoV for each channel; the dashed lines indicate the diffraction limit at the central wavelength.}
\end{figure}

The theoretical image quality of MAIA is excellent over the entire
FoV.  Figure~\ref{fig:ee80} shows the diameters that encircle 80\% of
the energy radiated by a point source, a criterion slightly more
strict than the full width at half maximum (FWHM), as a measure for
the width of the point spread function. Hardly any degradation of the
spatial resolution is expected over most of the FoV, down to the
Nyquist sampling limit of 0.55\,arcsec, defined by the plate scale of 0.276\,arcsec\,pixel$^{-1}$. Optical distortion
(pincushion) is small ($<$\,1\%) and, more importantly, very similar
for all three channels ($\pm$\,0.25\%). This ensures that small
identical windows can be used when reading out the same region of the
FoV from the three cameras. All lenses have broad-band anti-reflection
coatings with an average reflectivity below 1\% to increase
throughput. To ensure good $u$ efficiency, the collimator and the
\textit{u}\,camera only make use of glasses with very high UV
transmission.

The guiding camera is a commercial 35-mm reflex camera objective (Carl Zeiss Planar
T*\,1,4/50), with a focal length of \textit{f}\,=\,50\,mm and an aperture of
\textit{f}/1.4. This camera provides a FoV that is slightly larger than the science 
field on an ST-1603ME CCD camera (SBIG, USA) with 1020\,x\,1530 9-$\mu$m pixels. Unlike the
science cameras, the guiding detector does not have a field flattener. Hence,
image quality degrades substantially towards the edges of the field.

\subsection{Mechanical design}
\label{sec:mechanics}
\begin{figure}
\centering
\resizebox{\hsize}{!}{\includegraphics*{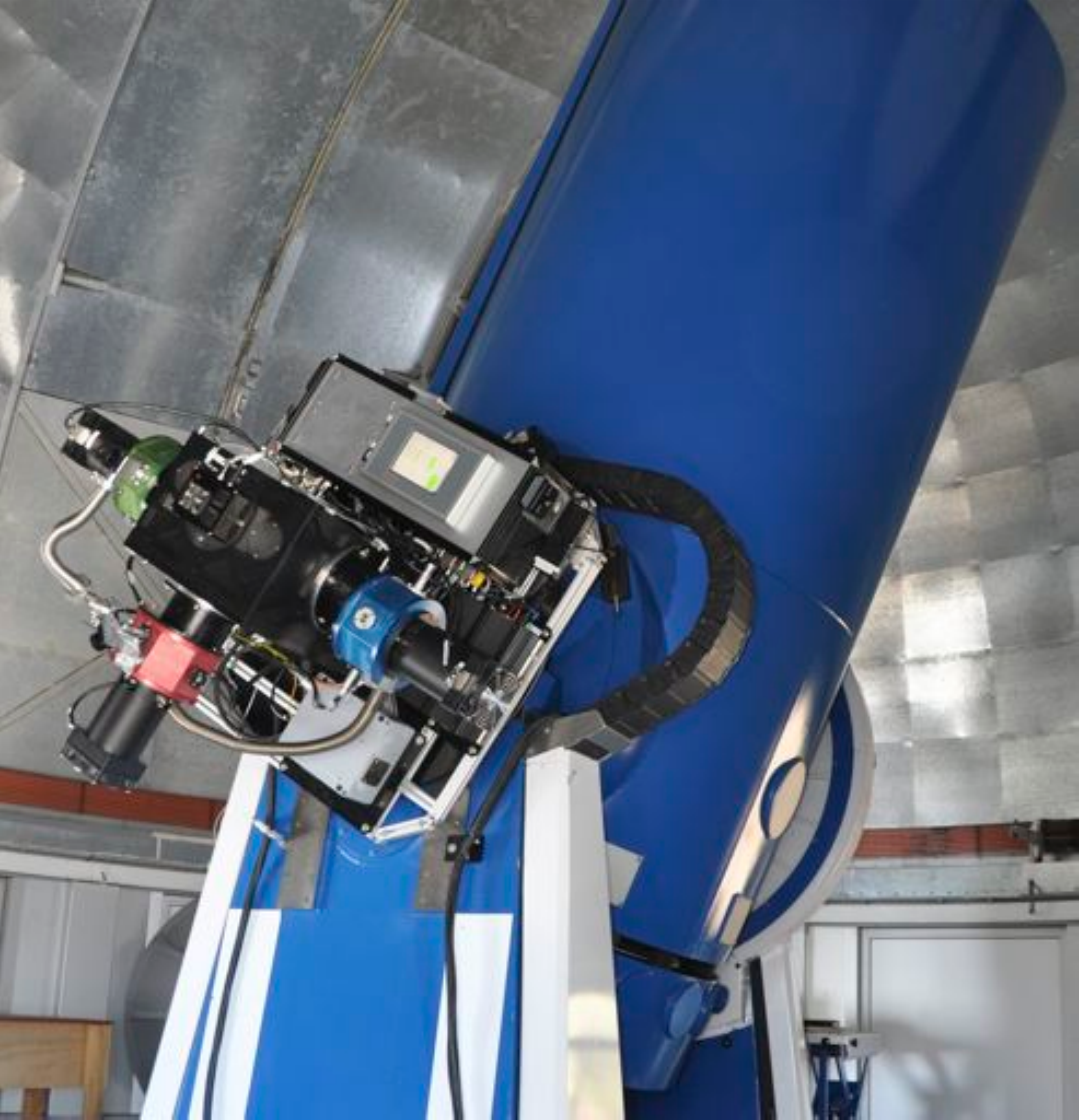}}
\caption{\label{fig:maiapicture}
Picture of MAIA mounted at the Nasmyth focus of the Mercator telescope. 
}
\end{figure}

\begin{figure*}
\centering
\resizebox{\hsize}{!}{\includegraphics*{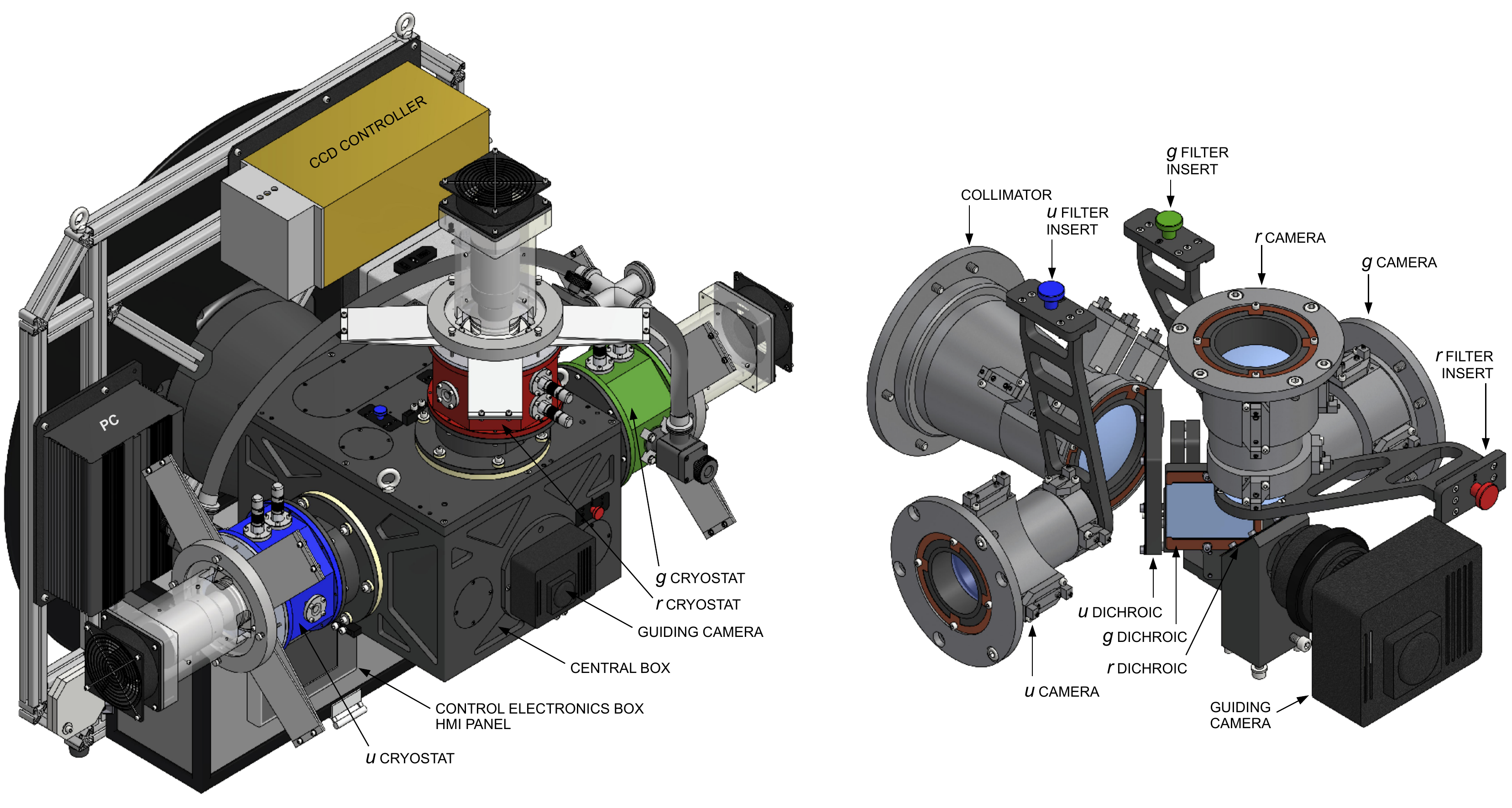}}
\caption{\label{fig:maia3d}
Left: 3D drawing of complete instrument, the colours of the three cryostats correspond to their respective  wavelength bands. Right: internal view of optical subsystems in the central aluminium box.
}
\end{figure*}

MAIA is installed on the instrument rotator at the Nasmyth\,B focal
station of the Mercator telescope (figure~\ref{fig:maiapicture}). The
3D drawings in figure~\ref{fig:maia3d} show the mechanical layout of
MAIA. The structural design of the instrument is based on a rigid
aluminium box that holds all the optical subsystems (cameras,
collimator, beam splitters). This ensures that the relative positions
of the optical elements are very stable. In any orientation, the
relative drift between the three channels is limited to just a few
detector pixels.  Homogeneous temperature changes will lead to isomorphic expansion of the aluminium box. This will not affect the angular positions of the cameras and the beam splitters with respect to the collimated beam, hence thermal miss-alignment between the three channels is avoided.
Precise machining of the interfaces to the optical
subsystems limits the number of required alignments. Access holes for
the adjustments of the beam splitter orientation are foreseen. The box
is extensively machined from a single block of cast aluminium,
reducing its mass to only 15\,kg.

The electronic accessories (CCD controller, data-acquisition PC, and 
control electronics box, housing the PLC with temperature control hardware, cryocooler drivers and power supplies) are not attached to the optics box but mounted on
a separate accessories frame that is directly attached to the
instrument rotator. This way, their weight does not load the
opto-mechanics of the instrument. An open cable reel is installed
between the accessories frame and the rotator. The total weight of
MAIA, accessories and cable reel included, amounts to 160\,kg. The
overall length of the instrument is 750\,mm and it rotates within a
660-mm radius.

\subsubsection{Lens mountings}
\label{sec:lensmountings}
The lenses of the collimator and the three cameras are mounted in four
lens barrels. The most critical parameter in the lens mountings is the
relative centring of the lens groups in their barrel. In order to
avoid image quality degradation, some of the lenses should be centred
with a precision of $\sim$0.03\,mm and this precision  needs to be
maintained over the operational temperature range (0$^{\circ}$C to
25$^{\circ}$C). Moreover, the instrument should be able to resist
storage and air transport conditions ($-$40$^{\circ}$C to
$+$55$^{\circ}$C).  Different coefficients of thermal expansion (CTE)
for the mounting barrel and the glass of the lenses prevents us from
limiting the radial play of the lens mounts to 0.03\,mm and hence, a
different centring concept is required.  Moreover, the optical glasses
of the lenses have very different CTEs, ranging from
0.55\,x\,10$^{-6}$ to 18.9\,x\,10$^{-6}$\,K$^{-1}$.

\begin{figure}
\centering
\resizebox{\hsize}{!}{\includegraphics*{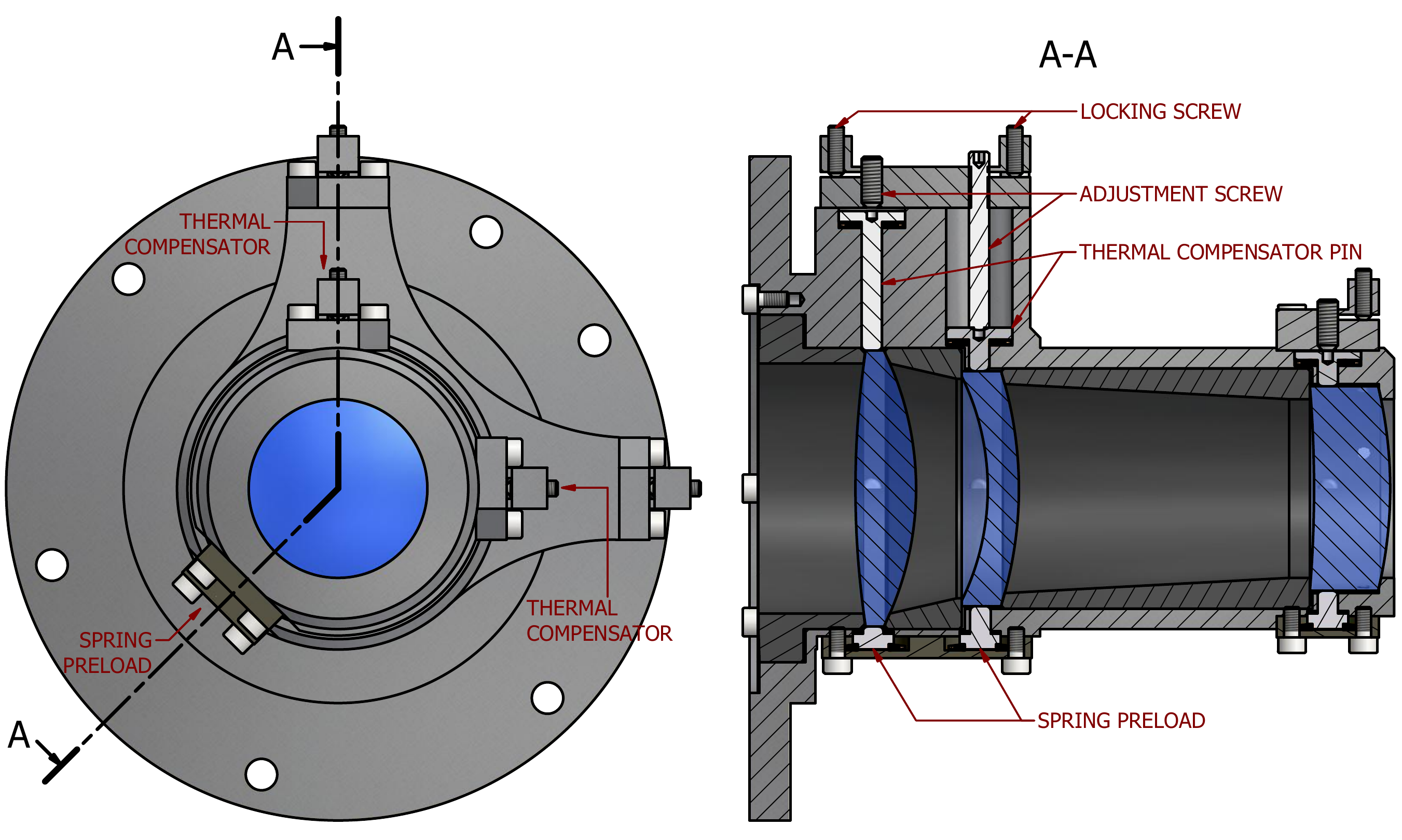}}
\caption{\label{fig:lensbarrel}
Front view and longitudinal section view of camera barrel; the thermal compensator pins and the spring preload constrain all lenses (blue) around the optical axis, located at the thermal centre of the system.
}
\end{figure}

To maintain the lenses centred over an extended temperature range,
their radial position is constrained by two pins (thermal compensator
pins) at a 90$^{\circ}$ angle (figure~\ref{fig:lensbarrel}). To align
the centring of the lenses, the effective length of these pins is
adjustable by lockable screws.  A third spring-loaded pin at
135$^{\circ}$ applies a constant force on the lens, pushing it towards
the centre while allowing for differential thermal expansion.  Both
the length and the material of the thermal compensator pins are chosen
in such a way that the thermal expansion of the glass, the lens barrel
and the pins, keep all lenses on the same optical axis over a broad
range of temperatures. We use aluminium, stainless steel or Invar as
materials for these pins, depending on the CTE of the lens
\citep{vandersteen12}.

\subsubsection{Cryocooler vibrations}
\label{sec:vibrations}
Free-piston Stirling coolers provide a compact and low-power alternative over
conventional liquid nitrogen or Joule-Thompsom (CryoTiger) cooling.  However, the
strong vibrations of the cooling engine can prohibit their application in
astronomical instrumentation. With the absence of a viable alternative (section~\ref{sect:cooling}), we decided nevertheless to
use Stirling cooling for MAIA. The piston of a Stirling cooler is driven by a linear
motor running at 60\,Hz. A resonating vibration absorber, precisely tuned to the
cooler frequency, is mounted on the backside of the cooler to absorb the bulk
of the harmonic disturbances. We measured that this absorber effectively reduces
vibrations at 60\,Hz to less than 10\% of their initial amplitude.

\begin{figure}
\centering
\resizebox{\hsize}{!}{\includegraphics*{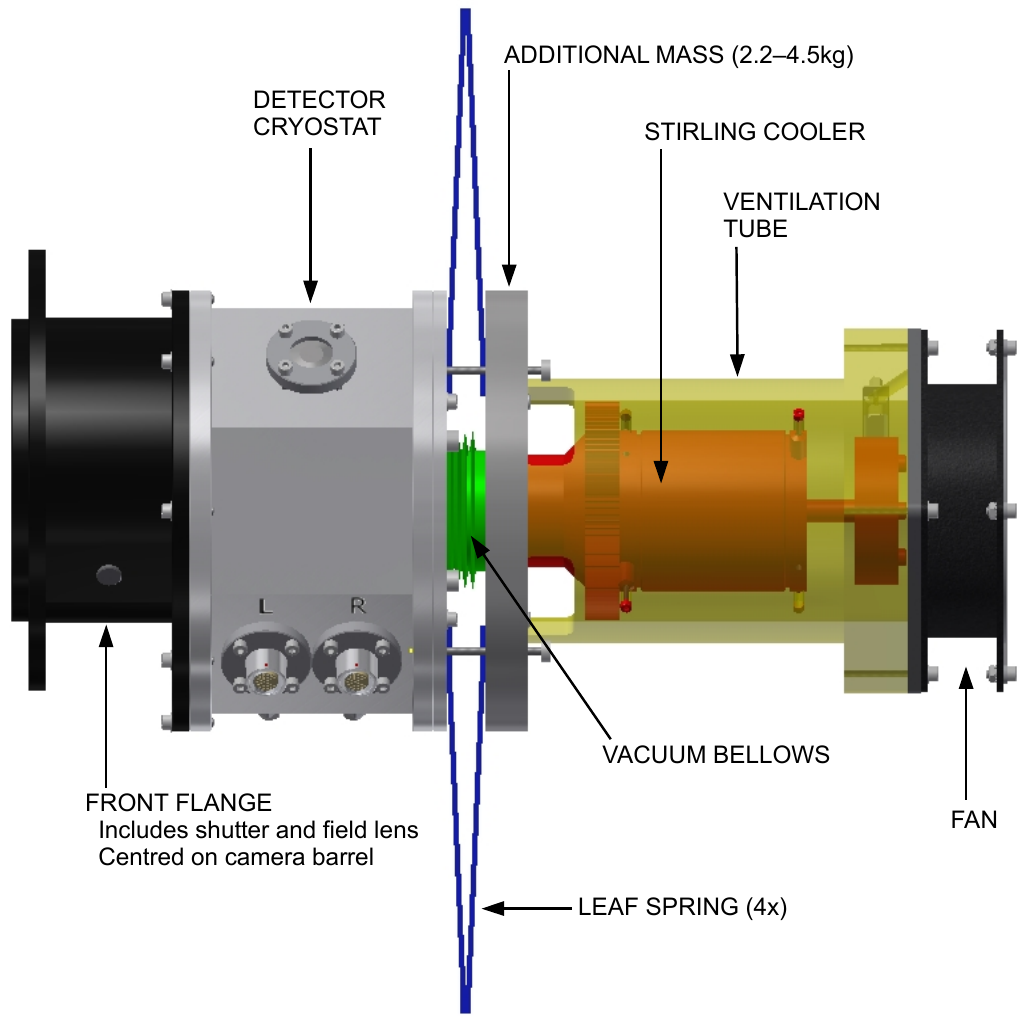}}
\caption{\label{fig:cryostat}
Drawing of a MAIA cryostat with spring-mounted Stirling cooler. The cooler is fixed in the ventilation tube (drawn transparently for clarity).
}
\end{figure}

\begin{figure}
\centering
\resizebox{\hsize}{!}{\includegraphics*{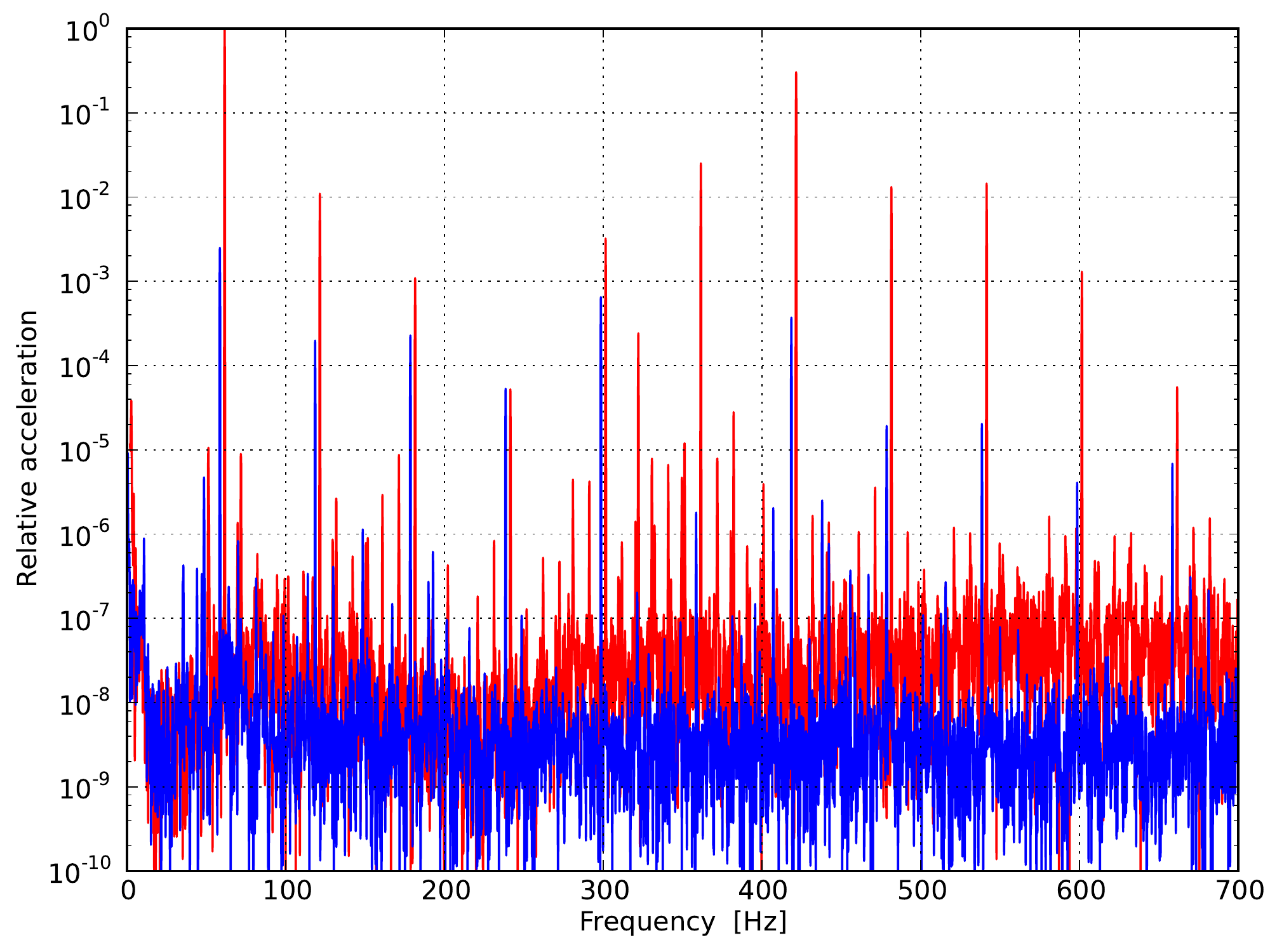}}
\caption{\label{fig:vibrations}
Instrument radial vibration spectrum without (red) and with (blue)
mass-spring decoupling between Stirling cooler and cryostat. 
Curves are translated by $+1$ and $-1$\,Hz for clarity.  Most
pronounced vibration peaks occur at multiples of 60\,Hz.}
\end{figure}

Laboratory measurements showed that the remaining cooler vibrations had no
detrimental effect on image movement or image quality. However, after
installation of MAIA on the telescope, significant vibrations of the
telescope structure appeared. These vibrations caused unacceptable oscillations
with amplitudes of up to 10\,arcsec. Spectral analysis of accelerometer
measurements on the instrument and the telescope showed that almost all
vibrational energy is contained within very narrow-band signals at 60\,Hz and
its higher harmonic frequencies (figure~\ref{fig:vibrations}). The narrow-band
nature of the telescope vibrations indicate that they are the result of direct
coupling to the Stirling cooler, rather than the excitation of a resonant mode
in the telescope structure. Although adding structural damping to this system
may intuitively be thought to attenuate vibrations, structural damping is more
effectively used to attenuate wide-band signals \citep{hartog84}. Therefore, we
mechanically decoupled the cooler from the rest of the instrument by 
mounting four leaf springs between cooler and cryostat (figure~\ref{fig:cryostat}).
A thin flexible bellows forms the vacuum interface between both parts. These springs  create
a low-pass mechanical filter that will no longer transmit vibrations of
frequencies that are substanstially higher than the eigenfrequency of the
system.

The eigenfrequency or resonance frequency $f_{\rm res}$ of a spring-mass system with
$m$ the mass of the cryocooler and $k$ the total spring constant of the leaf
springs and the vacuum  bellows, equals:

\begin{math}
\indent f_{\rm res} = \frac{\displaystyle 1}{\displaystyle 2\pi} 
             \sqrt{\frac{\displaystyle k}{\displaystyle m}}\,.
\end{math}
\vspace{5pt}

\noindent
To obtain a low eigenfrequency, a spring with small stiffness is
needed. However, the spring should be stiff enough to resist the force exerted by
the vacuum inside the cryostat ($\sim$160\,N at the altitude of the observatory) and to limit the axial and angular excursions of the cooler due to gravity when rotating the instrument. 
These constraints set a lower limit on $k$ of 33\,500\,N\,m$^{-1}$. To further reduce $f_{\rm res}$, we
increase $m$ by adding some extra mass to the cooler (initial weight:
$\sim$4\,kg). Each cooler receives a stainless steel mounting ring of different weight (2.2, 3.2 and 4.5\,kg for $r$, $u$ and $g$), resulting in different resonance frequencies for each cooler
(\mbox{$\sim$12}, 11, and 10\,Hz for $r$, $u$ and $g$), in order to avoid
resonant coupling between the three spring-mass systems.

Besides $m$ and $k$, the damping ratio $\zeta$ (a unitless measure that describes the oscillations' decay rate  after a disturbance) is a third parameter that characterises
the transmission at a given frequency. Large damping results in a small
resonance peak at the eigenfrequency but also in a reduced roll-off slope of the
low-pass filter and thus, maybe a bit counterintuitively, less efficient attenuation at higher frequencies. We did not encounter any resonance problems, so we tried to use as little damping as possible ($\zeta\simeq0.005$, a typical value for undamped metals).

To test the performance of the vibration isolation, we measured the
vibrational accelerations of the instrument and the telescope with and
without the spring isolators installed. Figure~\ref{fig:vibrations}
shows a typical vibration spectrum of the $r$ cooler (this is a worst
case because $r$ has the smallest extra mass and thus the highest
eigenfrequency). The accelerations were measured on the instrument
along the axis of the piston movement. Measurements at different
locations and in different directions give similar results but with
smaller amplitudes. The spring isolators reduce the harmonic peaks by
several orders of magnitude.  At the two main frequencies (60\,Hz and
420\,Hz), this reduction amounts to a factor of 500 and 800
respectively. While not really troublesome, broadband noise is also
reduced by an order of magnitude at higher frequencies. After the
installation of the vibration isolation, we could no longer detect any
perceivable effect on image quality when switching the coolers on or
off.

\section{Instrument performance}
\label{sec:performance}

\subsection{Efficiency}
To determine the throughput of MAIA on the Mercator telescope, we measured the photometric zero points for all three channels from a series of standard-star exposures at different air masses. These zero points correspond with the magnitude of a source that would give one detected photon per second outside the atmosphere. The results are given in table~\ref{tab:zeropoints}, together with the typical atmospheric extinction in each wavelength band. 
From these exposures, we also calculated the total efficiency of MAIA, excluding the telescope mirror reflectivity. It should be noted that these zeropoint measurements were performed with a telescope that had a five-year old aluminium mirror coating.  We estimated the mirror reflectivity at 83\% or less than 60\% for three reflections. A fresh layer of aluminium on all three telescope mirrors could increase the actual zero points by 0.25 magnitude.

In $g$ and $r$, the throughput corresponds with our expectations but not in $u$. This is mainly due to the low performance at short wavelengths of the anti-reflection coating on the detector. Since replacing the $u$ detector is not an option, we do not foresee an improvement for this in the near future. However, we also suspect that the transmission of one or more of the MAIA lenses might be  substantially  below specifications. We are still investigating this and hope to achieve a substantial increase of the $u$ throughput by replacing poorly performing optical elements.

\begin{table}
\centering
\caption{\label{tab:zeropoints}
MAIA measured and calculated peak efficiency (instrument and detectors without telescope), zero points (including telescope) and typical atmospheric extinction magnitudes at the Mercator telescope.}
\begin{tabular}{ccccc}
\hline\hline\noalign{\smallskip} \rule[0mm]{0mm}{1mm}
Band&  \multicolumn{2}{c}{Efficiency  (\%)} & Zeropoint & Atm. extinction \\
 & meas. & calc. & (mag) & (mag)\\
\hline\noalign{\smallskip}
\textbf{\textit{u}}  &   8.5  & 21 & 21.0 & 0.60 \\
\textbf{\textit{g}}  &  51 & 51& 24.0 & 0.12 \\
\textbf{\textit{r} } &  49 & 53& 23.7  & 0.09 \\
\hline
\end{tabular}
\end{table}

\subsection{Imaging performance}
Laboratory measurements of the image quality of MAIA without the telescope showed a point spread function with a FWHM of less than 17\,$\mu$m or 0.35\,arcsec over most of the FoV. This compares well with the theoretical maximum 80\% encircled energy diameters shown in Fig.~\ref{fig:ee80} These FWHM values should not degrade excellent seeing of 0.6\,arcsec by more than 0.1\,arcsec. As the collimator was optimised in combination with the Mercator telescope optics, performance at the telescope is expected to be at least as good. However, the first MAIA observing runs suffered from poor observing conditions and bad seeing. Moreover, the  overall image quality was further degraded  due to non-perfect alignment of the telescope optics.  As a result, image quality and spatial resolution have been severely limited up to now and hence, we could only prove that MAIA on the Mercator telescope is at least capable of producing 0.9-arcsec (FWHM) images in all three channels. The telescope misalignment is currently being corrected. Under more favourable atmospheric conditions, we expect to show substantially better imaging performance. 

All three cameras have to use the same window dimensions and location. To reduce read-out time, windows preferably are small, hence, it is important that the individual FoVs of the three cameras precisely coincide. The initial co-alignment of the three cameras is better than $\pm$\,5~pixels in the centre of the FoV. Due to different distortion by the optics of each camera ($\pm$\,0.25\%), an additional error of 8~pixels can occur at the furthest edge of the detector. Differential gravitational flexure when rotating the instrument causes an additional  drift of at most 3~pixels. Hence, a total shift of up to 16~pixels between the three cameras has to be taken into account when defining the window size.  Typical windows, depending on cycle-time requirements and assuming a worst-case co-alignment error and allowing for plenty of sky pixels, consist of 60\,--\,100 complete detector rows.

\subsection{First results}
\label{sec:results}
\begin{figure}
\centering
\resizebox{\hsize}{!}{\includegraphics*{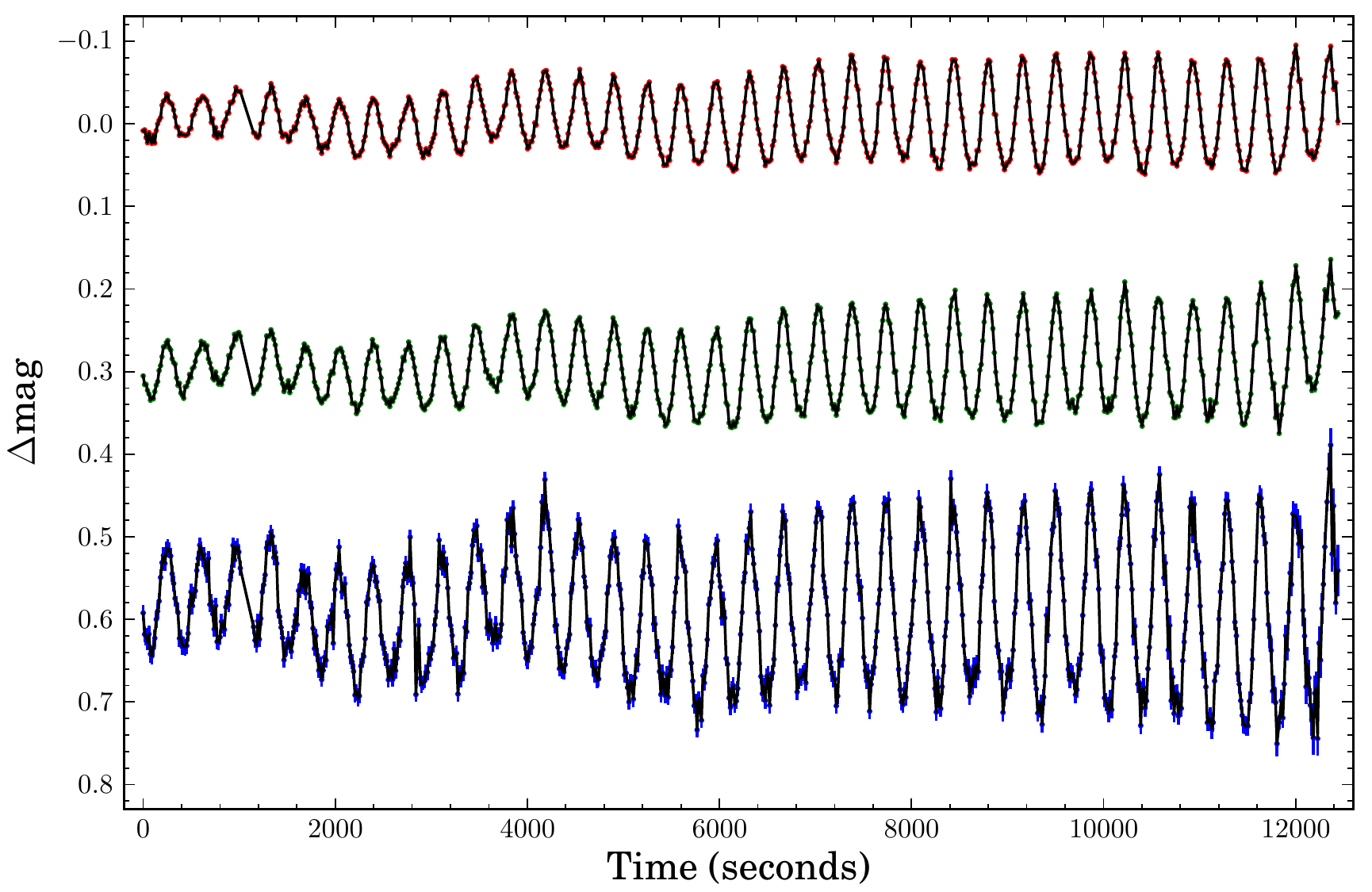}}
\caption{
MAIA light curve of Balloon taken with 15\,s integration times. The different colour bands (from top to bottom, $r$ in red, $g$ in green and $u$ in blue) are offset by 0.2 mag for clarity. The amplitude of the variations in $u$ is clearly higher than the variation in $g$ and $r$, as can be expected for a radial pulsation mode.
 \label{maia_FIG_bal09MAIA}}
\end{figure}

To illustrate the data that can be obtained with MAIA we show a 3.5\,h light curve of the high-amplitude hybrid subdwarf B star pulsator Balloon\,090100001 \citep[hereafter referred to as Balloon,][$V$\,=\,12.3]{OreiroUlla2004} in Fig.\ \ref{maia_FIG_bal09MAIA}.
The observations were performed during a commissioning run in October 2012 using 15\,s integration times. 

Balloon's highest amplitude pulsation mode at 2807.5\,$\mu$Hz (356\,s) is known to be the fundamental radial mode \citep{BaranPigulski2005}. It is clear from the figure that the pulsation amplitude is largest in the $u$, and smallest in the $r$ band, exactly as is expected for a radial $p$-mode \citep[see e.g.][]{RandallFontaine2005}. Unfortunately, the target has strong pulsation modes at 2823.2, 2824.8 and $2823.3\,\mu$Hz \citep{BaranPigulski2008}, which are close to the fundamental. Beating between these modes and the fundamental mode explains the reduced variability amplitude at the start of the MAIA observations. The dataset we obtained is not long enough to resolve the beating modes and hence does not allow us to reliably compare the pulsation amplitudes in the different bands. Detailed analyses of longer observation runs on other sdB pulsators will be presented in upcoming papers.

\section{Conclusions}
\label{sec:conclusions}

We have successfully built and commissioned MAIA, a three-channel imager capable
of producing high-speed three-colour light curves for variable star
research, with specific emphasis on pulsation mode identification capability for
asteroseismology.  With this instrument, we have put the detectors that were
developed for the Eddington space mission, to practical use in the science
field for which they were originally designed.  MAIA can increase the efficiency
of the Mercator telescope for multi-colour photometry by a factor of up to
six: a factor of three because of the simultaneous observations in three colour bands
and another factor of two thanks to the absence of dead time during detector read
out in case of fast-cadence observations. Furthermore, with a field of view of
9.4\,x\,14.1\,arcmin$^2$, MAIA has more than doubled the field coverage of the
Mercator telescope compared to the former MEROPE~II camera.

A limiting property of MAIA is the throughput in $u$ band that falls almost a
magnitude short with respect to the specifications, mainly due to the non-optimised
 anti-reflection coating of the detector. This non-compliance is still the subject of further study, with the aim of at least a
partial recovery of the $u$-band throughput in the future.

The combination of a dedicated telescope like Mercator and the MAIA instrument
offers unique capabilities for the study of variable stars, and in
particular for short-period pulsators.  The instrument will also
be used for follow-up observations of targets studied from MOST, CoRoT and {\it
  Kepler\/} space-based white-light photometry for which mode identification is
lacking and/or insufficient frequency precision occurs, preventing in-depth
seismic modelling. 

We plan to present the full details of the commisioning data
of a selected pulsating sdB star, along with a description of the data
reduction software, in a forthcoming paper. On completion of the
commissioning and software pipeline, MAIA will be offered to the
community on a collaborative basis keeping in mind the overall scheduling and
partnership requirements of the Mercator telescope.

\begin{acknowledgements}
This research was based on funding from the European Research Council under the
European Community's Seventh Framework Programme (FP7/2007--2013)/ERC grant
agreement n$^\circ$227224 (PROSPERITY) and from the Fund for Scientific Research
of Flanders (FWO), grant agreements G.0410.09 and G.0470.07, and the Big Science program. The CCDs of the MAIA camera were developed by e2v in the framework of the Eddington
space mission project and are owned by the European Space Agency; they were
offered on permanent loan to the Institute of Astronomy of KU\,Leuven, Belgium,
with the aim to build and exploit an instrument for asteroseismology research to
be installed at the 1.2m Mercator telescope at La Palma Observatory, Canary
Islands. Conny Aerts is grateful to Giuseppe Sarri and Fabio Favata for their
support and help in the practical implementation of the ESA loan agreement with
KU\,Leuven. 
We thank Simon Tulloch for his help in the optimisation of the detector control,
Tibor Ag\'{o}cs for his valuable comments on the optical design of MAIA and
Robin Lombaert for transporting the instrument to La Palma.
\end{acknowledgements}

\bibliographystyle{aa}

\end{document}